Four basic processes are envisioned, among them migration (diffusion), local rotation, isothermic chemical reaction, and nucleation. We center on nucleation for its being considered subject to classical physics. All of them are unified by a common approach to the barrier currents, that has been suggested as far back as 1961 by John Bardeen and then extended by Stefan Christov some ten years later. By introducing the respective radial potentials we incorporate Schrodinger's equation and thereby a quantal insight into the phenomena. Numerical calculations of the obtained statistical transition rates are reported.

1. Introduction

Barrier controlled processes are among the mechanisms playing an essential role in solid state physics. Despite recent advances many of them are still only poorly understood and need further studies, let alone the superconductivity in high-Tc cuprates or in iron based oxides [1,2]. Less known but not less important are the migration processes of impurities in the iron frame [3]. Lately, diffusion research has enriched knowledge of the iron system by obtaining solid evidence for a quantal diffusion of light carbon interstitials. However, the measured diffusion constants did provide neither the desired solid evidence for barrier heights nor for the configuration paths covered by the migrating impurity. This raised a few questions as to whether diffusion has not been interfered by another barrier controlled process making the identification impeded, carbon nucleation has been mentioned as a possibility. On the other hand, nucleation has recently joined the club of diffusion, local rotation and chemical reaction [4]. Indeed, nucleation *is* barrier-controlled (the work for forming a critical nucleus) and perhaps mode-coupled too.

The rates of chemical reactions [4], local rotation [5] and diffusion (migration) [3] are discussed at length elsewhere. In a unified rate form, the rates are composed as the weighted elementary contributions at the quantized vibronic energy levels $\nu W_n(E_n) Z_O^{-1} \exp(-E_n/k_B T)$, (vibrational frequency $\nu$ × the transition probability across the barrier $W_n(E_n)$ × the Boltzmann statistics factors), are summed up over the quantized energy levels $E_n$ on the left-hand side of the barrier. This implies that the quantized energy levels for the respective radial potential are available. In many cases the exact eigenvalues are *not* known and approximate substitutes (for instance, vibrational eigenvalues are used for the eigenenergies of a double-well potential). In another case, however, exact eigenvalues have been found and used for computing the transition probability – the nonlinear oscillator Mathieu functions which solved nearly exactly the local rotational problem [5].

In as much as we aim at substantiating the proposal for nucleation joining the club of mode-coupled and barrier-controlled rate processes, we shall carefully redefine the physical basis of our proposal so as to use approximations as little as possible. For this purpose the math problem will also have to be worked out on a respective solid basis. Again for the purpose of a comparative study we shall present illustrations of rates with various background to show just how they all work and compare with the statistical nucleation rate.

## 2. Bardeen's formula
### 2.1. Universal application

Generally, the transition probability $W_n(E_n)$ is defined in terms of the flux of vibrons in the initial electron state (left) along the reaction coordinate q towards the transition configuration at $q_C$. This flux is partially reflected back from the barrier and partially transmitted to the final electron state (right). The reverse current back from the final state may be neglected, if assumed that once in that state the vibron relaxes rapidly to lower levels giving away the excess energy through its coupling to the accepting modes, so that the chances for return are rather small. Under these conditions, the tunneling probability reads

$W(E_n) = j_{transmitted} / j_{incident}$

$$j(q) = ½ i \sqrt{(h\nu/M)} [\chi d\chi^*/dq - \chi^* d\chi/dq] \tag{1}$$

Undoubtedly, all the underlying quantities can be found by solving Schrödinger's equation with a radial potential. We presented the final results though the arguments involved are given below. They all follow suggestions originating from John Bardeen [6].

We have assumed that the motion along the configuration coordinate q is barrier controlled. In particular, the configurational transition probability along the radial coordinate based on the currents across the barrier will be [4]

$$W_{if\,conf}(E_n) = 4\pi^2 |V_{fi}|^2 \sigma_i(E_n)\sigma_f(E_n) \tag{2}$$

where the matrix element $V_{fi}$ is to be calculated using initial and final state wave functions $\phi_i$ and $\phi_f$, respectively, as ($\hbar = h/2\pi$):

$$V_{fi} = (-\hbar^2/2M) [\phi_f^* (d\phi_i/dq) - \phi_i (d\phi_f/dq)^*]|_{q=q_c} \tag{3}$$

Here $\sigma_i$ and $\sigma_f$ are the corresponding density-of-states (DOS) of the initial and final states. For a harmonic oscillator $\sigma_i(E_n) = \sigma_f(E_n) = (h\nu)^{-1}$. Inserting into (1) and performing the math in (2) we obtain the relevant formulas for the text.

Next, the respective Schrödinger equation reads

$$[(-\hbar^2/2M)\nabla^2\psi + \Delta V(r)\psi = E\psi \tag{4}$$

Equation (2) is our main proposal for it gives a simple way to buildup a quantum extension to topics traditionally considered classic. This may be the most simple way to incorporate quantum mechanics e.g. into nucleation. Otherwise, equation (4) may be considered in its own merit. Once its eigenvalue spectrum is available, we can construct a nucleation rate, classic or otherwise, by summing up the partial contributions of elementary rates at $E_n$:

$$\Re(T, r_C) = \nu (Z_A/Z_O) \Sigma_n W_n(E_n, r_C) \exp(-E_n/k_BT) \tag{5}$$

where $Z_A$ and $Z_O$ are the partition functions of the reactive modes and all the modes, respectively. In particular, $(1/Z_O) \exp(-E_n/k_BT)$ is the energy distribution function of Boltzmann's statistics which can be interchanged with any other statistics, if necessary. We also point out that the dependence of (5) on the barrier coordinate is through $W_n(E_n,r_C)$. Our upcoming problem is deriving appropriate expressions for the transition (tunneling) probabilities $W_n(E_n,r_C)$. Equation (5) is controlled by statistics, as shown elsewhere [7].

Equation (5) is *universal* even though originally worked out for the purpose of the reaction-rate theory. It gives an occurrence-probability approach to the rate of *any* mode-coupled barrier-controlled process. Different processes are distinguished by the specific probabilities at comparable vibronic energy levels.

### 2.2. Transition probabilities for double-well oscillators
#### 2.2.1. Application to chemical reactions, diffusion, as well as nucleation

In as much as the exact eigenvalue spectrum of the double-well oscillator is not available, approximations have frequently been used composed of linear combinations (gerade and ungerade) of the displaced-oscillator eigenstates pertaining to the bottoms of its left- hand and right- hand components [4].

Herein, we consider only strong-coupling configuration cases where the crossover coordinate is between the two well bottoms, left and right ones. For strong coupling situations we will assume the validity of Condon's approximation which factorizes out the electronic and configuration terms, as in:

$$W(E_n) = W_{el}(E_n)W_L(E_n) \qquad (6)$$

$W_{el}(E_n)$ is the probability (Landau-Zenner's) for a change of the electron state during the left-to-right transition across the barrier in Figure 1. $W_L(E_n)$ is the probability for configurational (lattice) tunneling across that same barrier.

$$\gamma(E_n) = (V_{12}/2h\nu) \sqrt{\{1/E_R|E_n-E_C|\}}$$

is Landau-Zenner's parameter. It has been obtained:

for overbarrier transitions at $E_n \gg E_C$:

$$W_{el}(E_n) = 2[1 - \exp(-2\pi\gamma)] / [2 - \exp(-2\pi\gamma)], \qquad (7)$$

for subbarrier transitions at $E_n \ll E_C$:

$$W_{el}(E_n) = 2\pi\gamma^{2\gamma-1} \exp(-2\gamma) / \gamma\Gamma(\gamma)^2$$

$$W_L(E_n) = \pi\{F_{nm}(\xi_{f0}, \xi_C)^2 / 2^{n+m} n!m!\}\exp(-[n-m]^2 h\nu/E_R)\exp(-E_R/h\nu) \qquad (8)$$

Here $Q = (n - m)h\nu$, the zero-point reaction heat, stands in the form of an energy conservation condition. n and m are the quantum numbers in the initial and final electron states, respectively. Further on

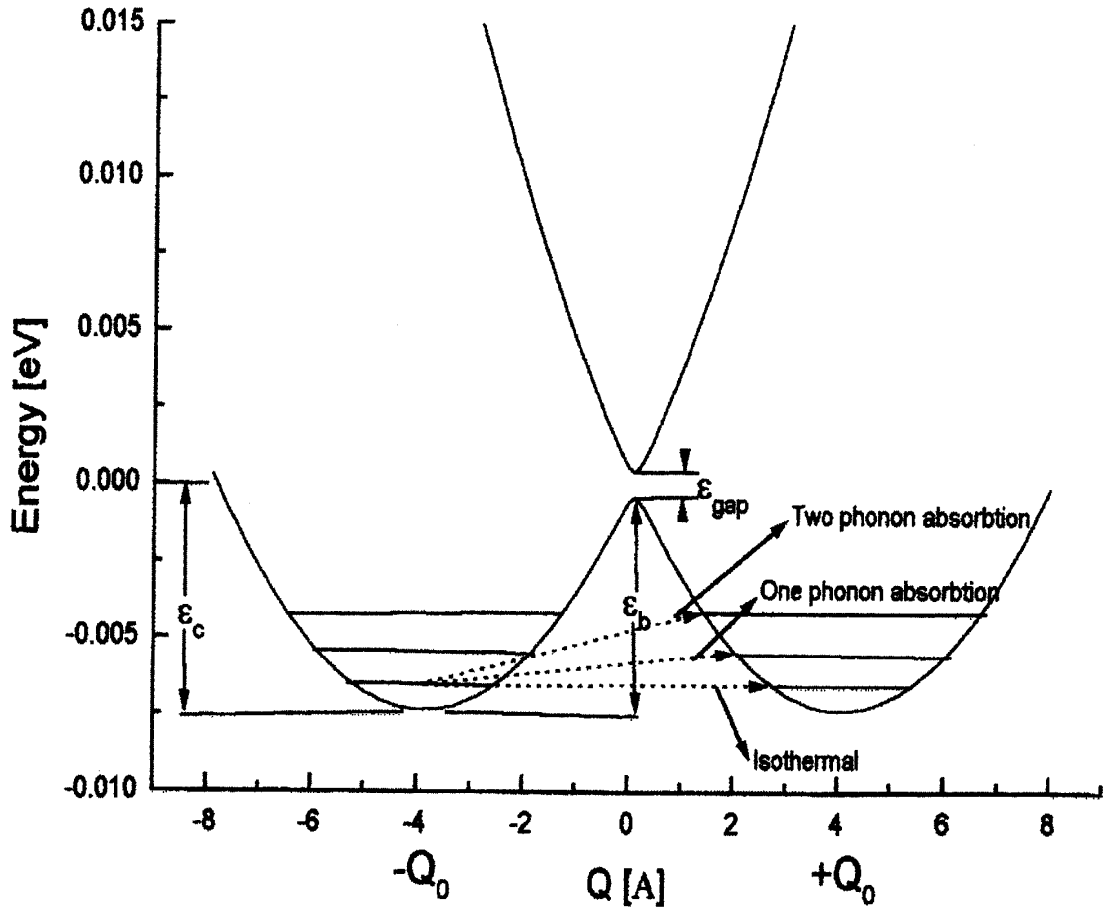

Figure 1: Sketch of the adiabatic potential energy surface (APES) involved in the vibronic transition from left to right: horizontal (0-phonon) tunneling, 1-phonon absorption, 2-phonon absorption, etc. The main part of the overall process is due to the 0-phonon horizontal tunneling with the casual participation of 1-phonon vertical tunneling. Both are easily identified in the temperature dependence of the rate in Figure 3. From Reference [3].

$$F_{nm}(\xi_{f0}, \xi_C) = \xi_0 H_n(\xi_C)H_m(\xi_C-\xi_0) - 2nH_{n-1}(\xi_C)H_{m-1}(\xi_C-\xi_0) + 2mH_n(\xi_C)H_m(\xi_C-\xi_0) \quad (9)$$

where $\xi = \sqrt{(M\omega^2/h\nu)}q$ is the dimensionless phonon coordinate, $\xi_{i0} = 0$ and $\xi_{f0} = \xi_0$ are the well-bottom phonon coordinates in the initial and final electronic states, respectively. Here and above $H_i(\xi)$ are Hermite polynomials of n-th order. The above equations hold good at $V_{12} \ll E_C = \Delta G(r_C) - \Delta G(r_O)$. This condition leaves blank a considerable portion of the energy axis.

A sketch pertinent to the equation (5) reaction is presented in Figure 1.

### 2.2.1.1. Transition probabilities for migration

The migration of a free atomic particle across a crystalline solid is both phonon-coupled and barrier-limited, due to the associated migration barriers. If the entity is a vibronic off-center polaron, then its classical trajectory should resemble a helix [8]. As a result, a migrating charge carrier may have a finite angular momentum just like a spinning particle. Quantum-mechanically, this is an azimuth in-plane motion in one of the off-center rotational bands combined with a plane-wave character along the perpendicular axis [9]. At this point we remind that a vibronic polaron is small and heavy just like an impurity atom. Unlike it, a diffusing impurity atom should cover a rectilinear path between two consecutive encounters with impeding barriers.

A diffusive motion of an impurity atom has been regarded as one coupled to a linear harmonic oscillator [10]. This oscillator is displaced by the amount determined by the coupling energy. For this reason it follows the steps of a reactive agent in an isothermic chemical reaction [11]. As to the alternative vibronic polaron, it has been represented as a free particle along the z-axis coupled to a nonlinear oscillator within the (x,y) plane [9]. The direction of propagation in this simplified case is the z-axis again. Figure 2 and 3 show the results of a migration study involving diffusion of carbon in iron [10] and interlayer or in-plane currents [16] in the $La_{2-x}Sr_xCuO_4$ superconductor, respectively.

### 2.2.1.2. Transition probabilities for nucleation

We shall spare some more space on this problem because of its somewhat controversial character. The nucleation within a solid matrix is usually tackled by means of the master rate equation [12]. Unlike it, we propose a statistical approach based on Bardeen's recipe. The nucleation-in-solid problem is that of the competition of surface and bulk free-energy terms [13]:

$$\Delta G(r) = 4\pi\sigma r^2 - (4/3)\pi r^3 \eta \ln(s/s_0) \tag{10}$$

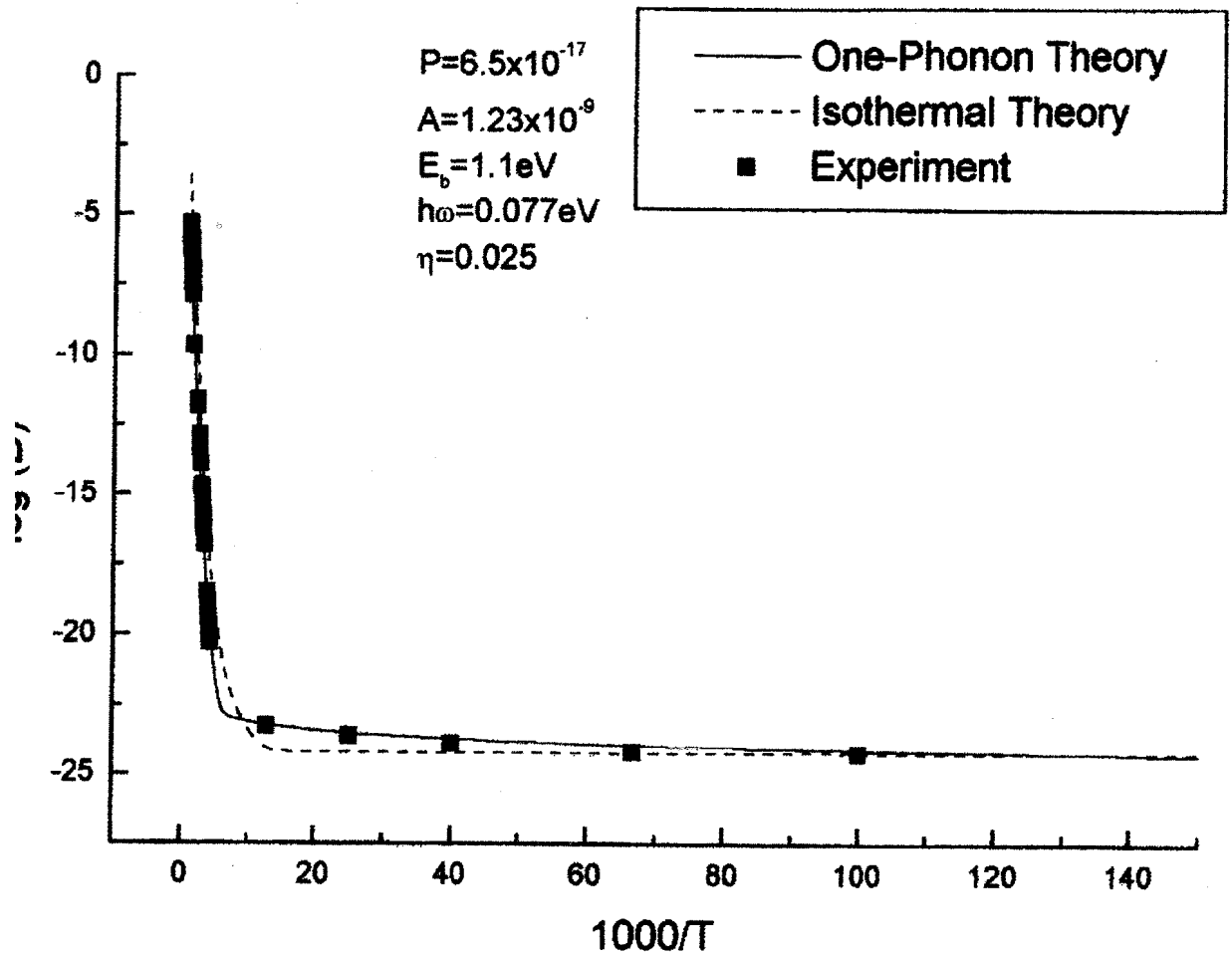

Figure 2: Temperature dependence of the *rate* as derived from diffusion data of C in α-iron from Reference [10]. There is also a comparison of the best fits to 0-phonon with 0-phonon + 1-phonon data which make the combined fit even more convincing.

where r stands for the cluster radius, σ is the specific surface energy, s = $s_0+\Delta s$ ($\Delta s>0$) is the supersaturation, $s_0$ being the saturation. η is the nucleation heat. By differentiating in r we obtain the extrema of (1) as follows: a minimum at r = 0, a maximum $\Delta G(r_C)$ at $r_C = 2\sigma / \eta\ln(s/s_0)$, and a possible second maximum beyond if the exhaustion of supersaturation due to cluster consumption is taken into account, not included in (1). $\Delta G(r_C) = (4\pi/3)\sigma r_C^2 = (4\pi/3)\sigma[2\sigma/\eta\ln(s/s_0)]^2$ is considered to be the work done for the formation of a critical nucleus, all nuclei at r < $r_C$ (embrios) being unstable to disintegration, the ones at r ≥ $r_C$ (nuclei) growing spontaneously to consume the supersaturation as nuclei of the daughter phase. Equation (4) is our main proposal for it gives a simple way to buildup a quantum extension to topics traditionally considered classic. Now, suppose equation (10) is at the same time the radial potential for the motion of a quantum mechanical particle along **r**, that is, Gibb's free energy is regarded as a potential energy for motion along r. The extension may be meaningful in so far as equation (10) is akin to an asymmetric double-well potential, as shown in Figure 1 of [4], provided the supersaturation consumption is taken for granted. Adherence to the reaction rate formula is the prerequisite for any reasonable statistical theory. If so, the ground-state radial Schrödinger equation reads

$$[(-\hbar^2/2M)\nabla^2\psi + \Delta G(r)\psi = E\psi \quad (11)$$

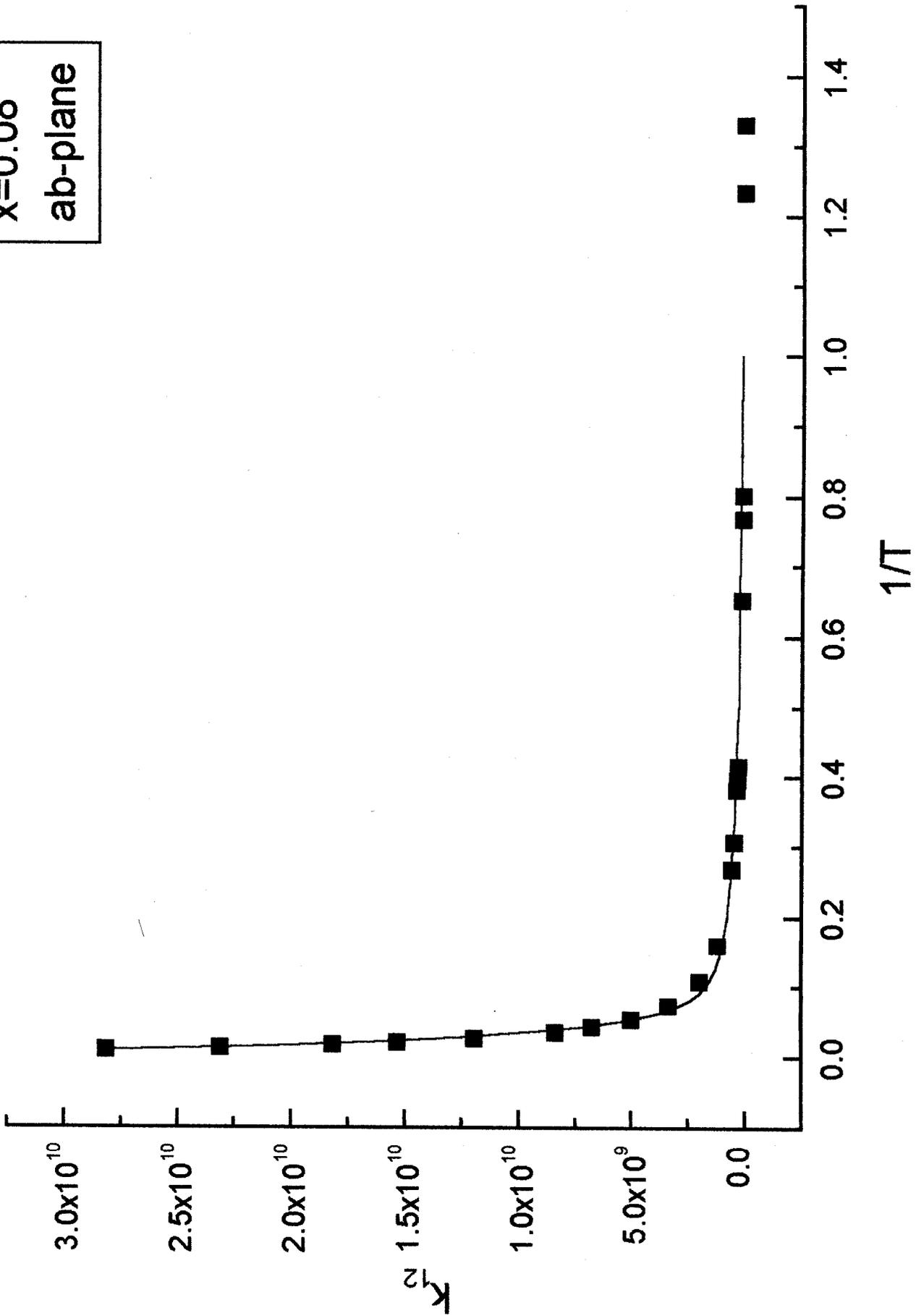

Figure 3: Temperature dependence of the *rate* as derived from data on in-plane currents in the $La_{2-x}Sr_xCuO_4$ superconductor from Reference [16]. An almost perfect fit has been obtained for lower doping concentrations (smaller x).

Equation (11) is our main proposal for it gives a simple way to build up a quantum extension to a topic traditionally considered classic. Perhaps it is not unique. Once its eigenvalue spectrum is available, we can construct a nucleation rate, classic or otherwise, by summing up the partial contributions of elementary rates at $E_n$, [14]:

$$\aleph(r,T) = \nu (Z_A/Z_O) \Sigma_n W_n(E_n,r_C) \exp(-E_n / k_B T) \quad (12)$$

Just to avoid useless polemics, we point out that the dependence of (12) on the barrier coordinate is in $W_n(E_n,r_C)$. Our further problem will be deriving the appropriate expressions for the transition (tunneling) probabilities $W_n(E_n,r_C)$. Equation (12) is controlled by statistics, as shown elsewhere [7].

### 2.3. Transition probabilities for nonlinear oscillators
#### 2.3.1. Application to local rotation and helical propagation

In order to construct initial and final states for Bardeen's formula we use Mathieu's periodic eigenfuctions and their derivatives. The basic problem was how best to describe the interior of rotational energy bands. Using notations standard for nonlinear oscillators, we get accordingly

$$W_{Ln}(E^n_m) = \begin{cases} 64 | (1-m)ce_{n-1}(z,q)[mdce_n(z,q)/dz] |^2_{z=\pi/2} (dm/da^n_m)^2 \\ 64 | (1-m)se_{n-1}(z,q)[mdse_n(z,q)/dz] |^2_{z=\pi/2} (dm/db^n_m)^2 \end{cases} \quad (13)$$

The above probabilities are maximum in the middle of a band at $m = \frac{1}{2}W_{Ln}^{max}$ and vanish at the band edges at m=0 and m=1. To work out an expression feasible for practical calculations the above equation should be normalized to 1. The normalized configuration probabilities are

$$W_{Ln}(E^n_m) = \begin{cases} 64N | (1-m)ce_{n-1}(z,q)[mdce_n(z,q)/dz] |^2_{z=\pi/2} (dm/da^n_m)^2 \\ 64N | (1-m)se_{n-1}(z,q)[mdse_n(z,q)/dz] |^2_{z=\pi/2} (dm/db^n_m)^2 \end{cases} \quad (14)$$

where the normalization factor is defined by

$$N^{-1} = 2\Sigma_{n=1}^{\infty} \int_0^1 W_{Ln}(E^n_m)dm = 128 \Sigma_{n=1}^{\infty} \int_0^1 dm[m(1-m)]^2 \times$$

$$\begin{cases} |ce_{n-1}(z,q)[dce_n(z,q)/dz]|^2_{z=\pi/2}(dm/da^n_m)^2 \\ |se_{n-1}(z,q)[dse_n(z,q)/dz]|^2_{z=\pi/2}(dm/db^n_m)^2 \end{cases} \quad (15)$$

In cases where Mathieu's functions can be approximated by free-rotor eigenstates $Y_m(\varphi,0) = \pi^{-1/2}\cos(m\varphi)$ we get $V_{fi}(E_n) \sim (h^2/2I\pi)m[-\cos[m(\varphi-\pi/4)]\sin[m(\varphi+\pi/4)] + \cos[m(\varphi+\pi/4)]\times\sin[m(\varphi+\pi/4)]]|_{\varphi=0}$ which is equal to $\pm(h^2/2I\pi)m$ for m odd and to 0 for m even. If we set $a_m = m^2$ leading to $\sigma(E_m) = (I/h^2)(1/m)$ we obtain $W_{if}(E_m) = 4\pi^2(h^2/2I\pi)^2 m^2 (I/h^2)^2(1/m)^2 = 1$ for m odd and $W_{if}(E_m) = 0$ for m even. It implies that the configurational probability of a free rotor is energy-independent, as it should. However if we set $a_m = a_m(q)$ leading to $\sigma(E_n) = (2I/h^2)[dn/da_n(q)]$ we obtain $W_{if}(E_m) = 4\pi^2(h^2n/2I\pi)^2 (2I/h^2)^2\}(dn/da_n)^2 = 4n^2 (dn/da_n)^2$ which is attributed to quasi-free rotations well above the barrier top.

For a hindered rotation, the eigenvalue spectrum being

$$E_{a/b,n} = (h^2/2I)a_n, b_n = (h^2/2I)[n^2 + c_n(q)], \qquad (16)$$

where $c_m(q)$ is the correction either to $a_m$ or to $b_m$ due to the hindering potential, we get

$Z_A/Z_O = 1/\Sigma_{m=0} \exp(-h^2[m^2+c_m(q)]/2Ik_BT)$

$= \exp(h^2c_0(q)/2Ik_BT) \times$

$\{1 + \Sigma_{m=1}^{\infty} \exp(-h^2[m^2+c_m(q)-c_0(q)]/2Ik_BT)\} \qquad (17)$

The graphic result for a local rotation *rate* is presented in Figure 4. For details see [17].

3. Conclusion

We believe to have shown beyond any doubt that the general multitude of formulas based on equations (1) through (5) of the text have an *universal* character in that they apply to *any* mode-coupled barrier-controlled process in a solid. They have already been applied quite successfully to the rates of chemical reactions, the rotation rates of nonlinear oscillators, as well as, quite recently, though, to the nucleation rate, an occurrence so deeply believed to belong to the category of purely classical problems.

Our quantal approach to nucleation can be applied to both the temperature-independent quantum-tunneling portion, with profound reference to quantum mechanical effects, and to a predominating strongly temperature-dependent barrier-controlled Arrhenius portion, where the effect of classical physics is predominating. It should be kept in mind, however, that in the overlapping portion both appearances can be identified. In the words of a leading solid-state British scientist, tunneling *is not* a *sole* property of low temperatures.

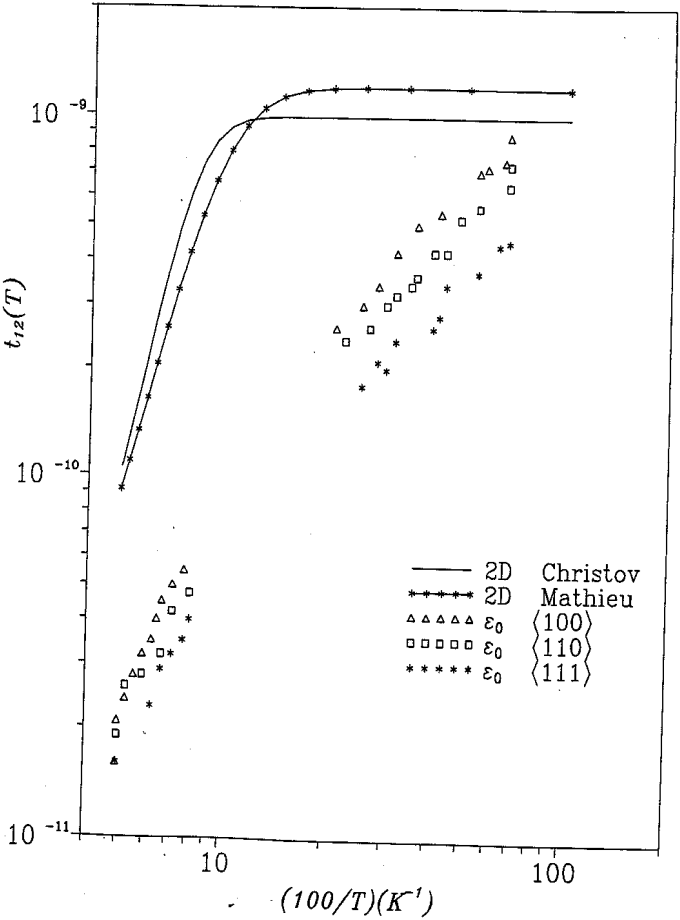

Figure 4: Comparison of the temperature dependences of the reciprocal rates, harmonic-phonon-coupled and the nonlinear-oscillator-coupled at the background of data on off-center $Li^+$ ions in KCl. The acute divergence between data and theory within the lower-temperature range is due to 1-phonon contribution, not accounted for at that time. From Reference [17].